\documentclass[twocolumn,amsmath,amssymb]{revtex4}
\usepackage{graphicx}
\usepackage{dcolumn}
\usepackage{bm}
\begin{document}

\title{Long Range Coulomb Interaction and the Majorana Fermions}
\author{Areg Ghazaryan and Tapash Chakraborty$^\ddag$}
\affiliation{Department of Physics and Astronomy,
University of Manitoba, Winnipeg, Canada R3T 2N2}

\date{\today}
\begin{abstract}
We have investigated the effects of long-range Coulomb interaction on the topological
superconducting phase in a quasi-one dimensional semiconductor wire, proximity coupled 
to a $s$-wave superconductor using the exact diagonalization approach. We find that in 
accordance with previous studies the addition of Coulomb interaction results in an
enlargement of the region of parameter values where topological superconductivity 
can be observed. However, we also find that although the interaction decreases the bulk 
gap for values of the magnetic field close to the phase transition point, for moderate 
magnetic fields away from the transition point, the interaction actually enhances the bulk 
gap which can be important for observation of topological superconductivity in this system. 
\end{abstract}

\maketitle

Majorana fermions (MFs) \cite{Majorana} have gained considerable attention in the 
past few years due to several proposals for the existence of Majorana modes in 
semiconductor systems \cite{Read,Fu,Oreg,Lutchyn,Alicea,Beenakker,Flensberg} as an 
elementary excitation. There were several experimental attempts to observe the MFs in 
semiconductor systems \cite{Mourik,Deng1,Das,Deng2,Churchill,Nadj-Perdge}. One of the 
most promising candidate for realization of the MFs \cite{Oreg,Lutchyn} is the observation 
of the topological superconducting phase in a one-dimensional (1D) semiconductor wire 
proximity coupled to a $s$-wave superconductor and with large Rashba spin-orbit (SO)
coupling \cite{Bychkov}. Considering a semiconductor system with high Land\'e $g$ factor and 
applying a magnetic field perpendicular to the Rashba SO coupling direction, the 
Kramers degeneracy can be lifted by inducing a large gap between the two states. By tuning 
the chemical potential of the system in the gap region, the system effectively becomes 
spinless and supports MFs at the edges of the wire similar to Kitaev's $p$-wave 
superconductor chain model \cite{KitaevModel}. It should be noted that the MFs can also 
be found in two-dimensional (2D) quantum wires where several subbands (channels) are 
occupied \cite{Potter,Fulga,Stanescu}, provided that the number of occupied subbands 
is odd and the width of the wire is smaller than the superconducting coherence length. 
Tilting of the magnetic field and the issue of orbital effects in 2D semiconductor wire 
have also been addressed \cite{Lim}.           

The major attraction for finding the MFs is due to their potential use in topological quantum 
computation \cite{NayakReview,SternReview}. The non-Abelian nature of the exchange statistics 
of the MFs makes fault-tolerant topological quantum computation feasible, which avoids the issue of 
decoherence in such a system. While for a strict 1D wire the exchange statistics is ill-defined, 
one can form networks of 1D quantum wires \cite{AliceaNetwork} and then move and exchange 
the MFs using closely spaced electronic gates. Therefore the quasi 1D wire systems where 
the MFs can be observed, has both theoretical and technological interest. 

Most of the recent studies have focused on the characteristics of topological 
superconductivity and the existence of the MFs in semiconductor wires without 
the interaction between the electrons in the wire. There were a few attempts 
to include electron-electron interaction in a strictly 1D wire using bosonization, 
density matrix renormalization and Hartree-Fock methods \cite{Stoudenmire,Gangadharaiah} 
and also in a 2D wire using the mean-field approach \cite{Manolescu}. The effect of 
the electron-electron interaction on charge jumps (which can be used to observe 
the MFs) in a 1D wire has also been discussed \cite{Shach}. It was observed 
that the interaction renormalizes the parameters of the system, namely, it enlarges 
the region of the chemical potential and the magnetic field strength, where 
topological superconductivity can be observed. For the 1D wire it was also found 
that interaction suppresses the bulk gap, and can eventually destroy it for high 
values of the interaction strength. However, the effect of the long-range Coulomb 
interaction on topological superconductivity has not been addressed as yet.

In this work we analyze the effect of long-range Coulomb interaction on the 
topological superconducting phase in a quasi 1D semiconductor wire, proximity 
coupled to a $s$-wave superconductor. In order to do that we employ the exact 
diagonalization scheme to obtain the energy eigenstates and the wave functions 
for the system with non-constant electron numbers with odd or even parity. We 
find that in accordance with previous studies, the addition of the Coulomb 
interaction results in an enlargement of the region of parameter values where 
topological superconductivity can be observed. We also find that although the 
interaction decreases the bulk gap for values of the magnetic field close to 
the phase transition point, for moderate magnetic fields away from the transition 
point it actually enhances the bulk gap, which can be important for observation 
of the topological superconductivity in this system. The exact diagonalization 
procedure employed here is quite general and can be used to address the issue 
of orbital effects and multichannel 2D quantum wire for observation of topological 
superconductivity in the presence of the Coulomb interaction between electrons. 

We consider a 2D semiconductor wire with hard wall confinement and a strong Rashba 
SO coupling \cite{Bychkov} in an applied magnetic field. As has 
been considered before, the semiconductor 1D wire is proximity coupled to a $s$-wave 
superconductor. We take the wire to be situated in the $xy$ plane, with $L^{}_x\gg L^{}_y$, 
where $L^{}_x$ and $L^{}_y$ are wire sizes in the $x$ and $y$ directions respectively. 
Without the superconducting pairing potential the Hamiltonian of the system is 
\begin{equation}
{\cal H}=\sum_{i}^{N^{}_e}{\cal H}_\mathrm{SP}^i+\frac12\sum_{i\neq j}^{N^{}_e}V^{}_{ij}.
\label{Ham2Dwire}
\end{equation}
Here $V^{}_{ij}=e^2/\epsilon\left|\mathbf{r}^{}_i-\mathbf{r}^{}_j\right|$ is the Coulomb 
interaction term and ${\cal H}^{}_\mathrm{SP}$ is the single-particle Hamiltonian, which 
can be written in the form
\begin{gather}
{\cal H}^{}_\mathrm{SP}={\cal H}^{}_\mathrm{W}+{\cal H}^{}_\mathrm{SO}+{\cal H}^{}_\mathrm{Z},\\
{\cal H}^{}_\mathrm{W}=\frac{\Pi_x^2+\Pi_y^2}{2m}-\mu+V^{}_\mathrm{C}(x,y),\\
{\cal H}^{}_\mathrm{SO}=\frac{\alpha}{\hbar}\left[\mathbf{\Pi}\times\boldsymbol{\sigma}\right]^{}_z, \\
{\cal H}^{}_\mathrm{Z}=\frac12g\mu^{}_\mathrm{B}\mathbf{B}\cdot\boldsymbol{\sigma}.
\end{gather}        
${\cal H}^{}_\mathrm{W}$ is the kinetic energy plus the confinement potential for 
the system with the chemical potential $\mu$, where $\mathbf{\Pi}=\mathbf{p}+(e/c)
\mathbf{A}$ is the canonical momentum, $\mathbf{A}=-B^{}_zy$ is the vector potential, 
$m$ is the effective mass. The confinement potential is $V(x,y)=0$, when $-L^{}_x/2<x<L^{}_x/2$ 
and $-L^{}_y/2<y<L^{}_y/2$, and $V(x,y)=\infty$ otherwise. ${\cal H}^{}_\mathrm{SO}$ 
is the Rashba SO interaction term, with the SO coupling strength $\alpha$. The Rashba 
SO coupling is considered to be present due to the confinement or the external electric 
field which create an asymmetry in the $z$ direction. Finally, ${\cal H}^{}_\mathrm{Z}$ 
is the Zeeman energy term, where $g$ is the Land\'e $g$ factor for the semiconductor. 
The magnetic field is taken to lie in the $xz$ plane with components $\mathbf{B}=
(B\sin\theta,0,B\cos\theta)$. By taking as the basis states the eigenstates of the ${\cal 
H}^{}_\mathrm{W}$ when $B=0$, namely
\begin{align}
\label{2Dwirebasis1}
\phi^{}_{n^{}_x}(x)=\sqrt{\frac{2}{L^{}_x}}\sin\left[\frac{n^{}_x\left(x+
L^{}_x/2\right)\pi}{L^{}_x}\right],\\
\phi^{}_{n^{}_y}(y)=\sqrt{\frac{2}{L^{}_y}}\sin\left[\frac{n^{}_y\left(y+
L^{}_y/2\right)\pi}{L^{}_y}\right],
\label{2Dwirebasis2}
\end{align}  
the Hamiltonian (\ref{Ham2Dwire}) can be cast into the second quantized form
\begin{align}
{\cal H}=\sum^{}_{nsn^\prime s^\prime}&\langle ns\left|{\cal H}^{}_\mathrm{SP}\right|n^\prime 
s^\prime\rangle c^\dagger_{ns}c^{}_{n^\prime s^\prime}+\nonumber \\
&\frac12\sum^{}_{\substack{n^{}_1s^{}_1n^{}_2s^{}_2 \\ n_1^\prime s_1^\prime n_2^\prime 
s_2^\prime}}\langle 
n^{}_1s^{}_1n^{}_2s^{}_2\left|V^{}_{12}\right|n_1^\prime s_1^\prime n_2^\prime s_2^\prime
\rangle \times \nonumber \\
 & \hspace{70pt} c^\dagger_{n^{}_1s^{}_1}c^\dagger_{n^{}_2s^{}_2}c^{}_{n_1^\prime s_1^\prime}
c^{}_{n_2^\prime s_2^\prime},
\end{align}
where for brevity $n=\{n^{}_x,n^{}_y\}$ has been introduced and $s$ denotes the spin quantum 
number of the particle. The proximity induced superconductivity potential can be written directly 
in the basis introduced above and has the form
\begin{equation}
{\cal H}^{}_\mathrm{SC}=\Delta\sum^{}_n\left(c^{}_{n\downarrow}c^{}_{n\uparrow}+c^\dagger_{n\uparrow}
c^\dagger_{n\downarrow}\right)
\end{equation} 
where the pairing potential strength $\Delta$ is taken to be real.

Instead of considering the Coulomb interaction at the mean field level \cite{Manolescu} where one
writes the complete Hamiltonian ${\cal H}^{}_\mathrm{PSC}={\cal H}+{\cal H}^{}_\mathrm{SC}$ in 
the Bogoliubov-de Gennes form, we directly treat the Coulomb interaction. In order to do that we use 
the exact diagonalization procedure to diagonalize ${\cal H}^{}_\mathrm{PSC}$ in even and odd sectors 
for small system sizes. For example, for the odd sector we diagonalize ${\cal H}^{}_\mathrm{PSC}$ 
for a system with non-constant number of electrons, namely $1,3,\dots N^{}_e$ electron number 
basis. A similar procedure is employed for the even sector as well. This gives us the possibility 
to obtain the low-lying energy states and the wave functions both for even and odd sector. This 
procedure can be used to investigate both the existence of the topological superconductivity 
and the effect of interaction on the pairing induced bulk gap. We use two complementary approaches 
to identify the topological superconducting phase described previously \cite{Stoudenmire}. 
If the electron number parity is a conserved quantity (which is the case for the Cooper pairing 
potential), then it is obvious even from the Kitaev model that the two degenerate ground states 
of the topological superconducting phase have different parity. Therefore the first notion of 
the topological superconductivity can be obtained by considering the value of 
\begin{equation}
\Delta E=\left|E^{}_\mathrm{odd}-E^{}_\mathrm{even}\right|,
\label{deltaE}
\end{equation}     
where $E^{}_\mathrm{odd}$ and $E^{}_\mathrm{even}$ are the ground states in the odd and even sector 
respectively. In ordinary superconductors the ground state is unique with integer number of 
Cooper pairs and therefore it has even parity. Hence $\Delta E$ is finite in ordinary 
superconductors. In the topological superconducting phase, when two MFs are exponentially 
localized at the two ends of the wire the $\Delta E$ is exponentially small. Therefore,
$\Delta E$ serves as some kind of order parameter to distinguish between the topologically 
trivial and non-trivial phases. 

In the second approach we calculate the Majorana wave functions from the obtained even 
and odd parity ground states. Let $|0\rangle$ and $|1\rangle$ be the even and odd parity 
ground states respectively. As was shown in the Kitaev model the odd sector state $|1\rangle$ 
is obtained by adding one non-local fermion, which is composed of two MFs localized at the two 
ends of the wire, to the even sector ground state $|0\rangle$, i.e., $|1\rangle=f^\dagger|0
\rangle$, where the fermion operator $f=\frac12(\gamma^{}_1+i\gamma^{}_2)$ and $\gamma^{}_1$, 
$\gamma^{}_2$ are the MF operators. Here the MF operators satisfy the well known relations 
$\gamma^2_a=1$ and $\{\gamma^{}_a,\gamma^{}_b\}=2\delta^{}_{ab}$. Using these operators 
we can write
\begin{equation}
|1\rangle=\gamma^{}_1|0\rangle=-i\gamma^{}_2|0\rangle.
\label{2Devenoddtr}
\end{equation}
Using the creation and annihilation operators $c^\dagger_{ns}$ and $c^{}_{ns}$ for the basis 
(\ref{2Dwirebasis1})-(\ref{2Dwirebasis2}), we can expand the MF operators $\gamma^{}_1$,$\gamma^{}_2$ 
in the form
\begin{equation}
\gamma^{}_a=\sum^{}_{ns}\left(\varphi^{(a)}_{ns}c^{}_{ns}+(\varphi^{(a)}_{ns})^\ast 
c^\dagger_{ns}\right)
\end{equation}
where $\varphi^{(a)}_{ns}$ are the expansion coefficients. By noting that $\{c^\dagger_{ns},
\gamma^{}_a\}=\varphi^{(a)}_{ns}$ and using (\ref{2Devenoddtr}) we get 
\begin{align}
\varphi^{(1)}_{ns}&=\langle0\left|c^\dagger_{ns}\right|1\rangle+\langle1\left|c^\dagger_{ns}
\right|0\rangle, \\
\varphi^{(2)}_{ns}&=-i\langle0\left|c^\dagger_{ns}\right|1\rangle+i\langle1\left|c^\dagger_{ns}
\right|0\rangle.
\label{2DwireMFscoeff}
\end{align} 
After obtaining numerically these expansion coefficients the probability distribution of the MFs
can be obtained via the relation $p^{(a)}(x,y)=\sum^{}_s\left|\sum^{}_{n}\varphi^{(a)}_{ns}
\phi^{}_{n^{}_x}(x) \phi^{}_{n^{}_y}(y)\right|^2$.

We now discuss our results for small number of electrons in a semiconductor wire 
proximity coupled to a $s$-wave superconductor using the exact diagonalization 
technique. The calculations were performed for the InAs semiconductor wire with 
following parameters: $m=0.042m^{}_0$, where $m^{}_0$ is the bare electron mass, 
$g=-14$, $\epsilon=14.6$ \cite{QDspinorbit}. The SO coupling strength was taken 
in all calculations to be $\alpha=45\,\mathrm{meV\cdot nm}$ and the superconducting 
pairing potential strength $\Delta=0.225$ meV. The wire sizes in all our calculations 
are $L^{}_x=3000$ nm and $L^{}_y=150$ nm. For the even sector we have considered 
up to eight electrons and seven electrons for the odd sector. This means that the 
number of electrons for the even sector in the many-body basis takes even values 
in the range from 0 to 8. Similarly for the odd sector it takes odd values in the 
range from 1 to 7. In order to achieve numerical convergence in our exact diagonalization 
calculations we have introduced a gap between the basis states $n^{}_x\leq7$ and 
$n^{}_x>7$. In a real system, this kind of gap can be achieved by adding a periodic 
potential to the system. The addition of the periodic potential can be regarded 
as a spatially varying chemical potential, and therefore it is desired that the 
periodic potential strength be smaller than the range of the chemical potential 
where topological superconductivity can be observed. Comparing this situation to the 
lattice model considered previoulsy \cite{Stoudenmire} it is natural to take 
$n^{}_x\leq N^{}_e$, where $N^{}_e$ is the maximum number of electrons in the 
system, so that every site is at least partially filled, which is the case in 
the density matrix renormalization group studies \cite{Stoudenmire}. In  this 
work we only discuss the $\theta=\pi/2$ case, which means that the magnetic 
field is aligned along the wire axis, and therefore we do not consider the orbital 
effects here. We also consider only the first transverse mode, i.e., we take 
$n^{}_y=1$ in all our calculation and therefore do not consider multi-channel effects. 
These two issues will be addressed in our future studies. 

\begin{figure}
\includegraphics[width=8cm]{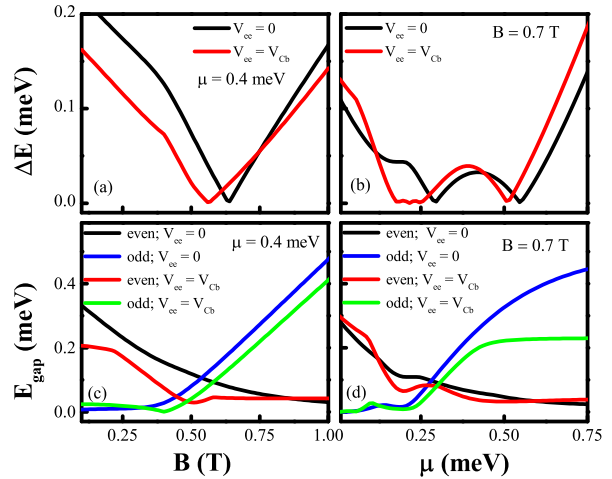}
\caption{\label{fig:En2dWireInt} (a) The dependence of the absolute difference 
between the energies of the ground states of odd and even sector on the magnetic 
field $B$. The chemical potential is taken to be $\mu=0.4$ meV, which corresponds 
to the transverse mode energy for $n^{}_y=1$. (b) Same as in (a) but for the 
dependence on the chemical potential for $B=0.7$ T. (c) The dependence of the bulk 
gap value (difference between the first excited and the ground state energy) for 
even and odd sector on the magnetic field $B$ for the chemical potential $\mu=0.4$ 
meV. (d)  Same as in (c) but for the dependence on the chemical potential for 
$B=0.7$ T.}\end{figure}

In Fig.~\ref{fig:En2dWireInt} (a) the dependence of the absolute value of the difference between the ground 
state energies in odd and even sector ($\Delta E$ defined in (\ref{deltaE})) on the magnetic field 
$B$ is shown for the chemical potential $\mu=0.4$ meV. Both cases of with and without the Coulomb 
interaction are presented. The value of $\mu=0.4$ meV corresponds to the energy of the
first transverse mode $n^{}_y=1$. For the quasi 1D wire the transition between the topologically trivial 
and non-trivial phases occurs when $V^{}_z=\sqrt{(\epsilon^{}_{n^{}_y}-\mu)^2+\Delta^2}$ \cite{Lim}, 
where $V^{}_z=g\mu^{}_\mathrm{B}B/2$ and $\epsilon^{}_{n^{}_y}=\hbar^2\pi^2n_y^2/2mL^2_y$ is the transverse 
mode energy. From Fig.~\ref{fig:En2dWireInt} (a) it is clearly seen that with an increase of the magnetic 
field strength, $\Delta E$ decreases for small values of the magnetic field, becomes zero at some 
point and then starts to increase. The zero point corresponds to a phase transition, 
where below that point the system is in the topologically trivial phase, whereas above that value 
it is in a topologically non-trivial phase. Due to the finite size effects $\Delta E$ is not exponentially 
small in the topological superconducting phase. For the non-interacting case the transition occurs 
at $B^{}_c=0.64$ T which is close to the value of $B^{}_c=0.56$ T for the ideal system calculated 
according to the relation above. Clearly, inclusion of the interaction results in the lowering of
the value of $B^{}_c$, which means that it extends the region of the magnetic field values where the
topological superconductivity can be observed \cite{Stoudenmire,Manolescu}. 

In Fig.~\ref{fig:En2dWireInt} (b) the dependence of $\Delta E$ on the chemical potential $\mu$ 
is shown for $B=0.7$ T. As can be seen here, $\Delta E$ becomes zero at two points, as expected, 
because for an idealized system the topological superconductivity should be observed in the range 
$\epsilon^{}_{n^{}_y}-\sqrt{V^2_z-\Delta^2}<\mu<\epsilon^{}_{n^{}_y}+\sqrt{V^2_z-\Delta^2}$. For an ideal 
system and for the values of parameters considered here we get the range $0.23\,\mathrm{meV}<
\mu<0.57\,\mathrm{meV}$, which is quite close to the values observed in Fig.~\ref{fig:En2dWireInt} 
(b). Similar to the case of Fig.~\ref{fig:En2dWireInt} (a) and also to the previous studies 
\cite{Stoudenmire,Manolescu}, we observe from Fig.~\ref{fig:En2dWireInt} (b) that inclusion 
of the interaction results in a broadening of the range of $\mu$ where topological 
superconductivity can be observed, except that in the present case the broadening is in the lower
values of the chemical potential.

In Fig.~\ref{fig:En2dWireInt} (c,d) the dependence of the energy difference between the first excited 
state and the ground state for both even and odd sectors on the magnetic field $B$ and the chemical potential 
$\mu$ is shown. As can be seen from these figures, in the topological superconducting phase the gap in 
the even sector is smaller than in the odd sector. Therefore this smaller gap corresponds to the bulk gap 
of the topological superconducting phase. In Fig.~\ref{fig:En2dWireInt} (c) 
this bulk gap decreases with increasing magnetic field for the non-interacting case. This is due 
to the fact that increasing the magnetic field aligns the spins of the electrons opposite to its 
direction, therefore reducing the pairing potential and destroying the superconducting phase. 
Inclusion of the interaction has two kinds of effects on the bulk gap. For magnetic fields close 
to the transition point value $B^{}_c$ the bulk gap is reduced. Therefore although the interaction broadens 
the region of the superconducting phase, it lowers the bulk gap, which makes the observation of 
the topological superconducting phase troublesome \cite{Stoudenmire}. Increasing the magnetic field 
strength further results in the dependence of bulk gap to become almost flat in case of the
interaction. In Fig.~\ref{fig:En2dWireInt} (c) for $B=1$T the bulk gap is 
already larger for the interacting case than that for the non-interacting case. As was shown previously 
\cite{Gangadharaiah,Stoudenmire} for even higher magnetic fields the interaction eventually 
suppresses the superconducting bulk gap. Therefore at the topological superconducting phase there 
is some middle region of the magnetic field values where the interaction favors the superconducting phase 
and enhances the bulk gap value. This finding can be important for possible observation of
topological superconductivity in this system. Finally, in Fig.~\ref{fig:En2dWireInt} (d) we see that for 
$B=0.7$ T, the interaction lowers the bulk gap for all values of $\mu$ where the topological 
superconductivity can be observed. 

\begin{figure}
\includegraphics[width=8cm]{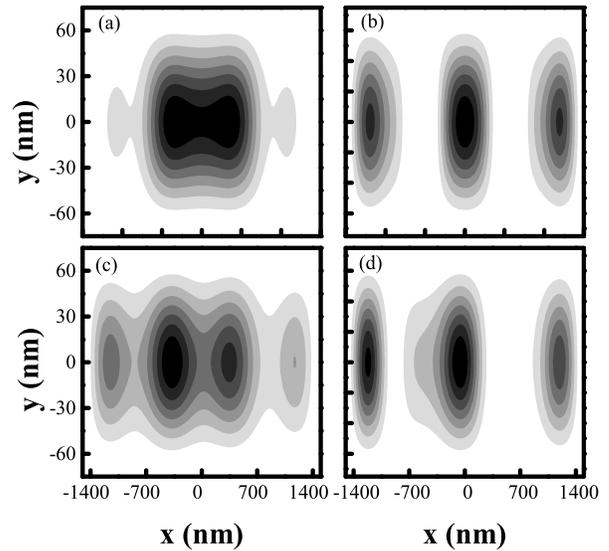}
\caption{\label{fig:Dens2DWireInt} The difference between the single-particle densities of the many-body 
states in odd and even sector without (a) and with (b) Coulomb interaction. Majorana fermion 
probability distribution for $\gamma^{}_1$ of (\ref{2Devenoddtr}) without (c) and with (d) 
the Coulomb interaction. For all figures, we take $\mu=0.4$ meV and $B=0.7$ T.}\end{figure}          

In Fig.~\ref{fig:Dens2DWireInt} (a,b) the difference between the single-particle densities of odd 
and even sector many-body state is shown without and with the inclusion of the Coulomb interaction. 
For all figures in Fig.~\ref{fig:Dens2DWireInt} the parameter values $\mu=0.4$ meV and $B=0.7$ 
T ($V^{}_z=0.28$ meV) have been used. In the ideal system the single-particle densities should be 
the same for odd and even sector \cite{Shach}, and therefore this density difference should be zero. 
In our calculation for the cumulative particle number difference through all the points of 
the wire and for the non-interacting case we get $\delta N=0.1$ particle difference which is also quite 
close to the value obtained previously for the 1D wire \cite{Shach}. When the Coulomb 
interaction is includede, this number decreases to the value $\delta N=0.03$. While it was predicted that 
for the non-interacting case the charge due to the MFs will have an oscillatory behavior in the wire 
and will be spread uniformly, as is shown in Fig.~\ref{fig:Dens2DWireInt} (a) the difference 
between the single-particle densities of the odd and even sector is not oscillatory and is mostly 
localized at the center of the wire. Further, adding the interaction effects has a considerable 
effect on this density difference, and now it is peaked both at the center and at the ends 
of the wire, despite the fact that it was predicted \cite{Shach} that the electron-electron interaction will 
not have a considerable effect on the charge distribution due to the MFs.

In Fig.~\ref{fig:Dens2DWireInt} (c, d) the MF probability distribution is shown for $\gamma^{}_1$ 
of (\ref{2Devenoddtr}) without and with the Coulomb interaction taken into account. The figures 
for the MF $\gamma^{}_2$ of (\ref{2Devenoddtr}) are inversion symmetric to the case of $\gamma^{}_1$ 
at the inversion point $x=y=0$. It should be noted that the procedure of obtaining the MF wave 
functions outlined above is exact when in the topological superconducting phase the ground state 
is doubly degenerate, which as we saw above is not satisfied for the realistic system. Therefore 
the MFs which we obtain in our calculation do not generally satisfy the $\gamma^2_a=1$ relation. 
This means that the MF wave functions obtained using the procedure above will only be 
approximately normalized for the system considered in this work. For the non-interacting case we 
get the normalization of the MF wave function to deviate from unity by about 4\%, whereas for 
the interacting case it is around 10\%. This difference between the non-interacting and interacting 
cases is related to the fact that for the interacting case the MF operator can have additional 
nonlinear terms in the expansion using the electron creation and annihilation operators 
$c^\dagger_{ns}$ and $c^{}_{ns}$ \cite{Stoudenmire}. As shown in Fig.~\ref{fig:Dens2DWireInt} 
(c) for the non-interacting case the MF probability distribution is shifted toward the left part 
of the wire, although its maximum is close to the center of the wire than to the edge. Adding 
interaction into the picture (Fig.~\ref{fig:Dens2DWireInt} (d)) splits the high peak into two 
parts, one located at the left edge and the other located at the center of the wire. Therefore 
the MF wave functions obtained with the procedure outlined in equations (\ref{2Devenoddtr}-
\ref{2DwireMFscoeff}) gives only approximate results in this exact diagonalization approach. 

In conclusion, we have considered the effect of long range Coulomb interaction on the
topological superconducting phase in a quasi 1D semiconductor wire proximity coupled to 
a $s$-wave superconductor using the exact diagonalization approach. By dealing with the problem 
using the exact diagonalization procedure we were able to treat the Coulomb interaction directly,
without resorting to any approximation. We have found that adding the interaction into 
the system enlarges the region of the parameter values, such as the magnetic field strength 
and the chemical potential, where topological superconductivity can be observed. This is consistent 
with the previous studies. In addition, we have also found that while for the magnetic field strength 
close to the phase transition point, the interaction lowers the bulk superconductivity gap, for
moderate values of the magnetic field strength away from the transition point, the interaction 
actually enhances the bulk gap. This finding can be important for observation of the topological 
superconducting phase in the experiment due to the fact that the effect of the Coulomb interaction 
can be controlled by manipulating the charge density in the semiconductor wire. Finally, the 
present approach is quite general and can also be used to consider the properties of
topological superconductivity in a 2D semiconductor wire with long-range Coulomb interaction,
and include additional features such as the orbital effects or the multichannel filling. 

The work has been supported by the Canada Research Chairs Program of the Government
of Canada.

\end{document}